\documentclass[%
 amsmath,amssymb,
 aps,
 pre,
 reprint,%
]{revtex4-1}

\usepackage{graphicx}
\usepackage{dcolumn}
\usepackage{bm}
\usepackage[all]{xy}
\usepackage{subfigure}
\usepackage{sidecap}

\begin{document}


\title{Transition from homogeneous to inhomogeneous steady states in oscillators under cyclic coupling}
\author{Bidesh K. Bera$^{1}$}
\author{Chittaranjan Hens$^{2} $}
\author{Sourav K. Bhowmick$^{3} $}
\author{Pinaki Pal$^{4} $}
\author{Dibakar Ghosh$^{1}$}
\email{diba.ghosh@gmail.com}

\affiliation{$^{1}$Physics and Applied Mathematics Unit, Indian Statistical Institute, Kolkata-700108, India\\
$^{2}$Department of Mathematics, Bar-Ilan University, Ramat Gan, Israel\\
${3}$ Department of Electronics, Asutosh College, Kolkata 700 026, India\\
${4}$ Department of Mathematics, National Institute of Technology, Durgapur 713209, India.
}%

\date{\today}

\begin{abstract}
We report a transition from homogeneous steady state to inhomogeneous steady state in coupled oscillators, both limit cycle and chaotic, under cyclic coupling and diffusive coupling as well when an asymmetry is introduced in terms of a negative parameter mismatch. Such a transition appears in limit cycle systems via pitchfork bifurcation as usual. Especially, when we focus on chaotic systems, the transition follows a transcritical bifurcation for cyclic coupling  while it is a pitchfork bifurcation for the conventional diffusive coupling.  We use the paradigmatic Van der Pol oscillator as the limit cycle system and a Sprott system as a chaotic system. We verified our results analytically for cyclic coupling and numerically check all results including diffusive coupling for both the limit cycle and chaotic systems.

\end{abstract}

\maketitle

A variety of self-organized behaviors  appears in coupled dynamical units due to diverse forms of interactions and connectivity.
Quenching of oscillation \cite{Awadesh-rep, Aneta-rep} is one such example of  natural phenomena, which exists  and
controls  the dynamics of coupled oscillatory systems to establish a desired state which is significant in the perspectives of biological systems \cite{Koseska-synthe,diverse_route_epl}, and also chemical  oscillators \cite{yosh}. It is established that, in diffusively coupled systems, if the mismatch in parameters is large enough, the oscillation may be suppressed and the systems are stabilized to a fixed point \cite{Kopell} or multiple fixed points \cite{Bar-eli}.
This phenomenon may also occur if a sufficient time delay in the interaction \cite{Sen} among identical or mismatched dynamical units is present. Other mechanisms of suppressing oscillation are related to  dynamic  coupling \cite{Konishi}, environment coupling \cite{Resmi}, conjugate coupling \cite{Rajat} or repulsive interaction \cite{Hens} which are relevant for different other contexts.
Recently such an emergence of stable fixed points has been precisely distinguished into two broad classes \cite{Aneta-rep,Hens,Volkov, Kurths-pre, Banerjee, Banerjee1, Hens2}: amplitude death (AD) and oscillation death (OD). AD  denotes a suppression of oscillation to a single or homogeneous steady state (HSS) which is preferably linked to stabilization of the uncoupled system's equilibrium. OD is defined as inhomogeneous steady states (IHSS) when all the oscillators populate different stable fixed points  which have  dependency on the strength of interaction. The HSS is manifested as a stabilization or suppression of oscillation in neuronal system \cite{Ermentrout-Siam}, while the IHSS is connected to the cellular differentiation \cite{Kaneko} and also in synthetic genetic networks \cite{Koseska-synthe}.  The HSS and the IHSS may also coexist  in a multicellular population \cite{Koseska}. We  mention here that coexistence of OD and limit cycle has  also been experimentally demonstrated and numerically verified in coupled Van der Pol system \cite{Volkov-IJBC}.

\par An issue of prime importance is a possible transition from HSS to IHSS which is very much likely to occur in dynamical systems, in general, physical, chemical, ecological and biological. Long back, Turing \cite{Turing} reported a symmetry breaking diffusion process that creates an instability to induce a transition in a homogeneous medium  to stable patterns. A similar transition of a stable HSS state to stable IHSS states was first discovered, recently, in a simple model of two Landau-Stuart (LS) oscillators for diffusive coupling \cite{Volkov, Aneta-rep} due to an asymmetry created by parameter mismatch. This encourages a lot of interest to confirm this phenomenon of HSS-IHSS transition for a variety of other natural forms of coupling, namely, delay coupling \cite {Kurths-pre}, conjugate type and dynamic coupling \cite{Kurths-pre}, and repulsive interaction \cite{Hens} and a weighted mean-field coupling \cite{Banerjee, Banerjee1, Pooja-pramana} and linear augmentation \cite{Pooja-pre}. In limit cycle systems, it is so far found that the transition is always, irrespective of the type of coupling and source of symmetry breaking, occurs via pitchfork bifurcation (PB) which is similar to the Turing \cite {Turing} bifurcation. 
Comparatively, the chaotic oscillators have not been given much attention to investigate such a transition. In recent past, the HSS-IHSS transition was investigated  in chaotic oscillators under both attractive diffusive and a repulsive mean-field perturbation \cite{Hens, Hens2} and it is found to follow diverse routes, transcritical bifurcation (TB) or saddle-node bifurcation (SNB). To the best of our knowledge, the HSS-IHSS transition has not been explored so far in chaotic oscillators under diffusive coupling or any other coupling form except the repulsive interaction. 
\begin{figure}[h]
 \centerline{\includegraphics[height=5.0cm,width=6cm]{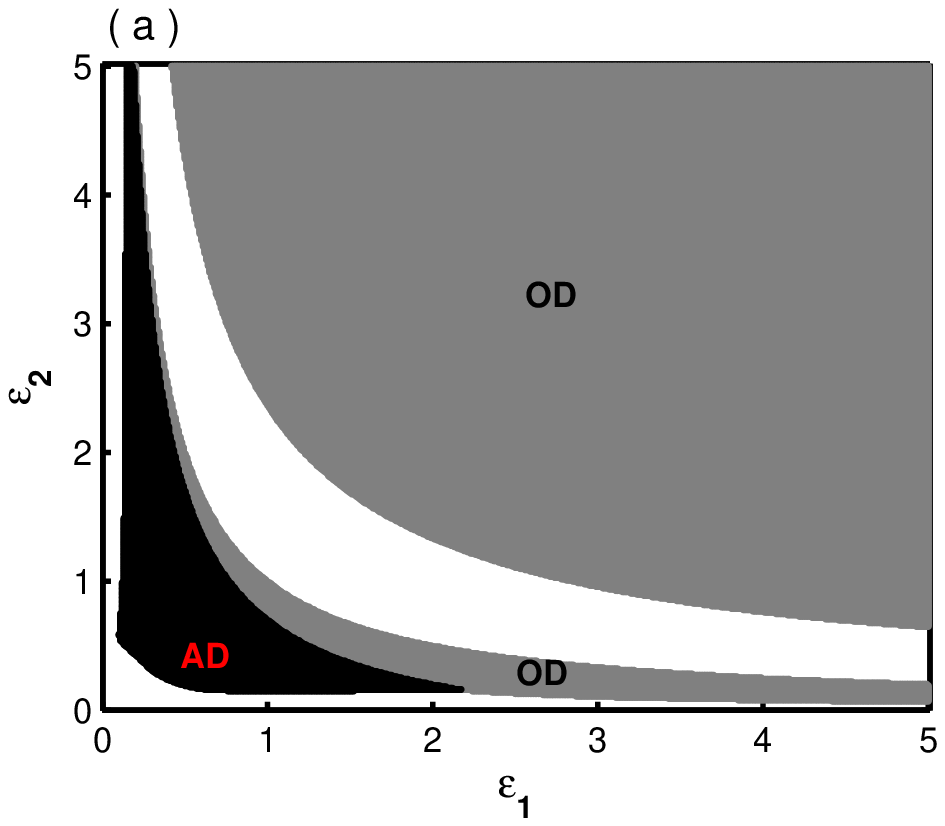}}
 \centerline{\includegraphics[height=5.0cm,width=6cm]{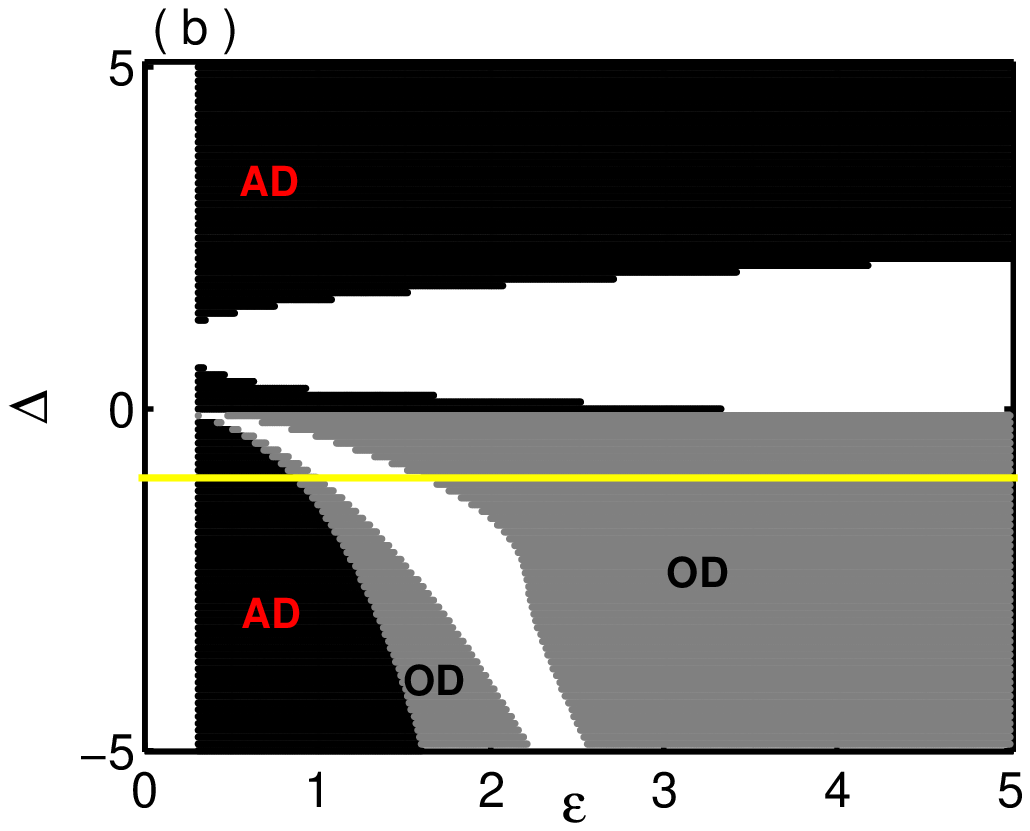}}
 \centerline{\includegraphics[height=5.0cm,width=6cm]{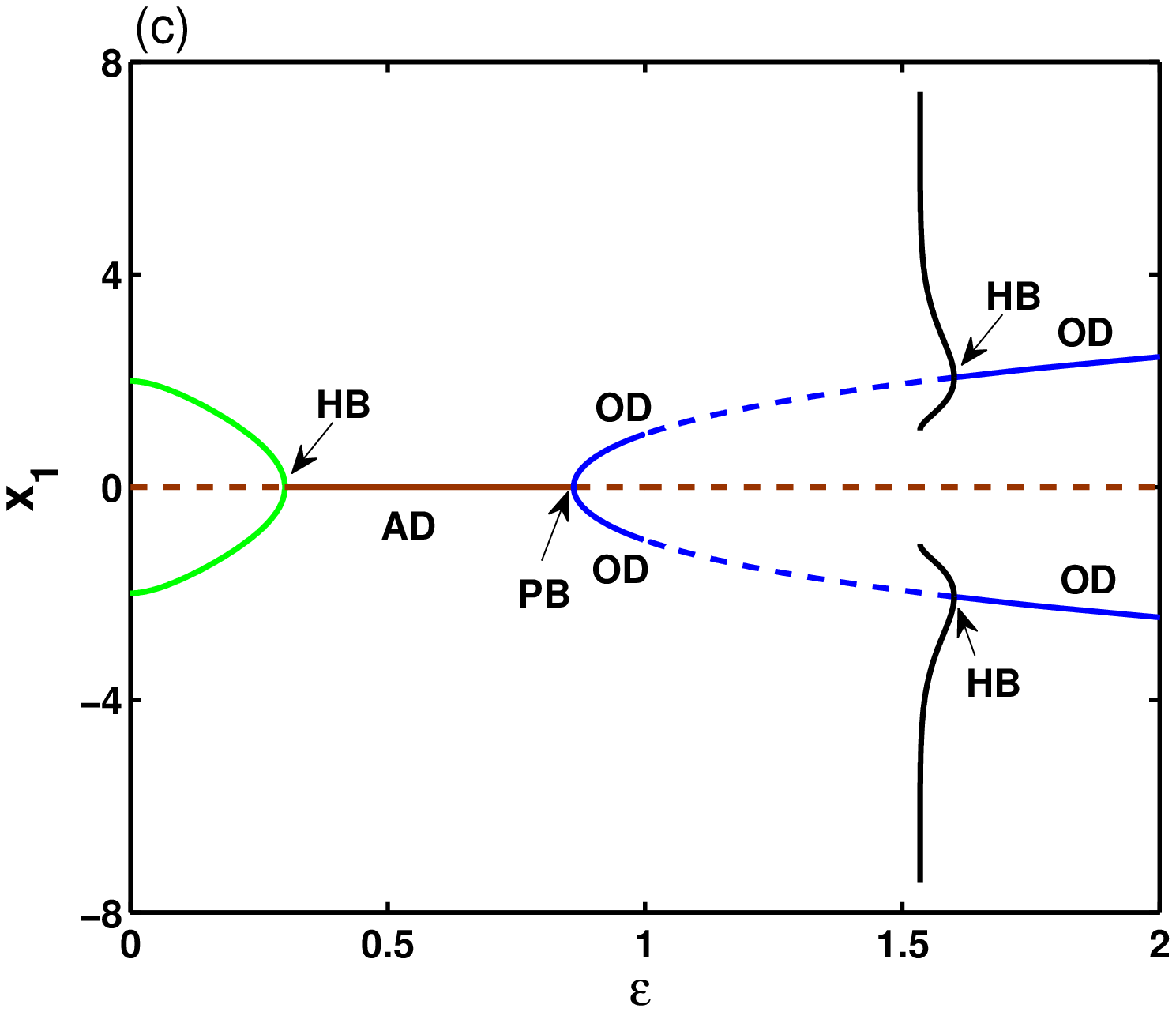}}
 \caption{Coupled Van der Pol oscillators under cyclic coupling.  HSS (AD) and IHSS (OD) islands in (a) ($\epsilon_1$-$\epsilon_2$)  plane, (b) ($\Delta$ - $\epsilon$) plane.  Black  and gray regions represent AD and OD respectively. White region represents the coupled system is in oscillatory state. (c) Bifurcation diagram of cyclic coupling (taking parameters along the diagonal line of \ref{epsilon_12}(a) or the yellow line of \ref{epsilon_12}(b)) shows a transition from HSS to IHSS where extreme values of $x_{1}$ are plotted with coupling strength $\epsilon$. The coupled system undergoes transition from LC to AD at $\epsilon=0.3$ followed by a transition to OD at $\epsilon=0.86$.  Further increase of $\epsilon$ another OD regime appears via a inverse Hopf Bifurcation(HB).}
\label{epsilon_12}
 \end{figure}
\par In this perspective, we use here a cyclic coupling scheme \cite{Dana_epjst} in limit cycle systems as well as chaotic systems and try to understand the process of HSS-IHSS transition under a parameter mismatch. 
A cyclic coupling is generated by two separate interaction: a unidirectional diffusive coupling link which connects two oscillators in a forward direction via a pair of variables while another unidirectional link connects the oscillators in a reverse direction via a second pair of variables and thereby maintain a mutual interaction between the oscillations. 
The cyclic coupling thus defines a diffusion process or mutual exchange of information but do not involve the same pair of state variables of the oscillators. In the perspectives of engineering, under the cyclic coupling, one oscillator sends a signal via one pair of variables to the other oscillator while receives a feedback via another pair of variables \cite{Dana_epjst}. In biological systems, this is similar to neuronal interactions where one neuron sends a signal via one pair of synaptic dendrites and receives a feedback via another pair \cite{Kandel}.
We first investigate two Van der Pol (VDP) oscillators under cyclic coupling. We find out a region of parameter mismatch where the HSS-IHSS transition occurs  via the typical PB route.

Two coupled  VDP oscillators under cyclic coupling are, 
\begin{eqnarray}
\label{vdp1st}
&&\dot x_{1}=\omega_{1}y_{1}+\epsilon_{1}(x_{2}-x_{1})\nonumber\\
&&\dot y_{1}=b(1-x_{1}^{2})y_{1}-\omega_{1}x_{1}\nonumber\\
&&\dot x_{2}=\omega_{2}y_{2}\\
&&\dot y_{2}=b(1-x_{2}^{2})y_{2}-\omega_{2}x_{2}+\epsilon_{2}(y_{1}-y_{2})\nonumber
\end{eqnarray}
The first oscillator receives a signal  from the second oscillator via the $x$-variable and sends a feedback via the $y$-variable which is expressed as cyclic coupling. $\epsilon_{1,2}$ is the coupling strength, $\omega_1$ and $\omega_2$ are the mismatch parameters and $b$=0.3. If we couple in reverse order i.e, the first oscillator is coupled with $\ y$ variable and second oscillator is coupled with $\ x$ variable the behavior of coupled systems and transition from HSS to IHSS are same. The coupled VDP system has a trivial fixed point $(0,0,0,0)$ which is the HSS solution of the system and the other fixed points (IHSS solutions) are $(x_{1}^{*},\:y_{1}^{*},\frac{\epsilon_{2} y_{1}^{*}}{\omega_{2}},0)$ where $y_{1}^{*}=\frac{\omega_{1}x_{1}^{*}}{b(1-x_{1}^{*2})}$ and $x_{1}^{*}=\pm\sqrt{1-\frac{\omega_{1}^{2}\omega_{2}+\epsilon_{1}\epsilon_{2}\omega_{1}}{b\omega_{2}\epsilon_{1}}}$.
\\ Jacobian matrix at nonzero fixed point $(x_1^*, y_1^*, x_2^*, y_2^*)$ is given by
 $$J=\left( \begin{array}{cccc}
  -\epsilon & \omega_{1} & \epsilon & 0 \\
                    p_{1} &p_{2} & 0 & 0 \\
                    0 & 0 & 0 & \omega_{2} \\
                    0 & \epsilon &p_{3} &p_{4}

\end{array} \right)$$

where $ p_{1}=-\omega_{1}-2bx_{1}^{*}y_{1}^{*}, p_{2}=b(1-x_{1}^{*2}), p_{3}=-\omega_{2}-2bx_{2}^{*}y_{2}^{*}$ and $p_{4}=b(1-x_{2}^{*2})-\epsilon.$ The characteristic equation of the above Jacobian matrix is
\begin{eqnarray}
&&\lambda ^{4}+a_{1}\lambda ^{3}+a_{2}\lambda ^{2}+a_{3}\lambda +a_{4}=0
\end{eqnarray}

where $a_{1}=\epsilon -p_{2}-p_{4}, \;\; a_{2}=-\epsilon p_{2}-\omega _{1}p_{1}-p_{3}\omega {2}-p_{4}\epsilon +p_{2}p_{4}, \;\; a_{3}=p_{2}p_{4}\epsilon +\omega_{1}p_{1}p_{4}-p_{3}\omega_{2}\epsilon+p_{2}p_{3}\omega_{2}, \;\; a_{4}=p_{2}p_{3}\omega_{2}\epsilon +p_{1}p_{3}\omega_{1}\omega_{2}-\epsilon^{2}p_{1}\omega_{2}.$ \\

We check the stability of the equilibrium points as a function of $\epsilon_1$ and $\epsilon_2$ in Fig. \ref{epsilon_12}(a) when $\omega_1=-\omega_2=1$. The trivial fixed point origin, i.e, the HSS(AD) is stable in the black region whereas the nontrivial fixed points, i.e, the IHSS are stable in the gray regions. The boundary between  HSS(AD) and IHSS(OD) is given by the curve $\omega_{2}\epsilon_{1} b=\omega_{1}^{2}\omega_{2}+\epsilon_{1}\epsilon_{2}\omega_{1}$.
The oscillatory states are represented by the narrow white regime of extreme left and in the extreme down. Interestingly, two sets of nontrivial fixed points exist, one group is stable in the upper gray region and the other set is stable in the lower gray region when we have taken, $\omega_1=1$ and $\omega_2=-1$ and  the unstable island is  shown between them as white regime.
We explore the HSS and IHSS scenarios in a broader parameter space of $\Delta$- $\epsilon$ plane assuming $\epsilon_1=\epsilon_2=\epsilon$. Note that, for HSS and its transition to IHSS, a symmetry breaking parameter is necessary while the cyclic coupling is still maintaining a symmetry so far $\epsilon_1=\epsilon_2=\epsilon$. A parameter mismatch $\Delta=\frac{\omega_2}{\omega_1}$ introduces the essential asymmetry in the coupled system. We fix $\omega_1$ at +1 and vary $\omega_2$ continuously from -5 to +5 with a step size of 0.1 and draw the  $\Delta$ vs $\epsilon$ plot in Fig. \ref{epsilon_12}(b).  For a positive mismatch, $\Delta>0$, the limit cycle (LC) eventually collapses to stable HSS. Further increase of coupling strength cannot  induce IHSS in the coupled systems. On the other hand  when $\Delta \sim 0$, a small HSS island exists since $\omega_2$ is too small and the uncoupled second oscillator which contains $\omega_2$ behaves like bursting. Obviously, for $\Delta\sim 1$, no death island appears due to a small or no mismatch of the coupled systems. However, distinct regions of LC, HSS and a direct transition from HSS to IHSS is clearly seen in the $\Delta<0$ region when $\omega_2$ has a negative value  and it signifies a change of orientation of the trajectory or a counter-rotation \cite{ Bhowmick, Witkowski, Prasad, Newby} of the Van der Pol system. Two oppositely rotating VDP oscillators show a HSS to IHSS transition when they are coupled via a type of cyclic coupling.
Thus the negative parameter mismatch has a strong bearing on the HSS to IHSS transition as seen in the Fig. \ref{epsilon_12}.
\par We derive the stability condition of the trivial fixed point origin assuming $\epsilon_1=\epsilon_2=\epsilon$ and $\Delta=-1$. The stability of the equilibrium origin is calculated  from the {\it Jacobian} of the coupled VDP,
$$J=\left( \begin{array}{cccc}
  -\epsilon & \omega_{1} & \epsilon & 0 \\
                    -\omega_{1} & b & 0 & 0 \\
                    0 & 0 & 0 & \omega_{2} \\
                    0 & \epsilon & -\omega_{2} & b-\epsilon

\end{array} \right)$$

with eigen values $\lambda=\frac{1}{2}[(b-\epsilon)\pm\sqrt{(b-\epsilon)^{2}+4k}]$
where $k=-\frac{1}{2}[-(1-b\epsilon)\pm\sqrt{4\epsilon^{2}+\epsilon^{2}b^{2}}]$.
The stability condition of the fixed point origin is obtained as
$b\le\epsilon\le\frac{1}{2}(-b+\sqrt{b^{2}+4})$.
For a choice of $b=0.3$, HSS occurs in the coupling range $0.3\le\epsilon\le0.86$ and the IHSS appears at $\epsilon=0.86$ which match with the numerical simulations along the horizontal line in Fig. \ref{epsilon_12} (b).
\par Further we have checked the nature of HSS to IHSS transition using  the MATCONT tool \cite{Dhooge} and plotted the extrema of $x_1$  with coupling strength $\epsilon$  in Fig.\ref{epsilon_12}(c) for $b=0.3$ and $\omega_1=-\omega_2=1$. For a coupling strength $\epsilon<0.3$ the coupled system shows limit cycle behavior. At $\epsilon=0.3$, the coupled system stabilizes to the HSS via Hopf bifurcation (HB) and it continues until $\epsilon=0.86$ when two stable nonzero equilibrium points i.e. IHSS appear via PB and continues up to 1. It is to be noted that if  $\omega_1=-\omega_2=1$ and   $\epsilon=1$,  $x_1^*$ becomes 1 and $ y_1^*$ becomes undefined which creates a singularity at $\epsilon=1$. With further increase of $\epsilon$ ($\epsilon>1$) an unstable limit cycle appears which eventually goes to a stable limit cycle (The extrema  is shown by black line) with high amplitude at $\epsilon=~1.6$. Further increase of $\epsilon$ generates another IHSS becomes stable via inverse HB and is shown by  solid blue line in Fig.\ref{epsilon_12}(c).
The MATCONT plot basically confirmed the analytical prediction whereas  the nature of transition matches with Fig \ref{epsilon_12}(a) along diagonal line or the yellow line of the Fig\ref{epsilon_12}(b).
\par An immediate question arises how the coupled VDP oscillator would behave if the negative mismatch ($\Delta=-1$) is introduced under the typical diffusive coupling instead of the cyclic coupling? 
 For this we consider two diffusively coupled VDP systems,
\begin{eqnarray}
&&\dot x_{1}=\omega_{1}y_{1}+\epsilon_1(x_{2}-x_{1})\nonumber\\
&&\dot y_{1}=b(1-x_{1}^{2})y_{1}-\omega_{1}x_{1}+\epsilon_2(y_{2}-y_{1})\nonumber\\
&&\dot x_{2}=\omega_{2}y_{2}+\epsilon_1(x_{1}-x_{2})\nonumber\\
&&\dot y_{2}=b(1-x_{2}^{2})y_{2}-\omega_{2}x_{2}+\epsilon_2(y_{1}-y_{2})
\label{vdp2nd}
\end{eqnarray}
 Consider two types of  diffusive coupling through one variable i.e, (i) $\epsilon_{1}=0, \epsilon_{2}\neq 0$ or (ii) $\epsilon_{1}\neq 0, \epsilon_{2}=0$.  In the first type of coupling $x_{1}-x_{2}$ will be in anti-synchronization state and $y_{1}-y_{2}$ are in complete synchronization state i.e, mixed synchronization is occurred \cite{Bhowmick, Bhowmick2} for $\omega_2=- \omega_1$. Again $x_{1}-x_{2}$ are in complete synchronization and $y_{1}-y_{2}$ are in anti synchronization state for the second type of coupling. Other possible choice of couplings are explained in details in Ref \cite{Bhowmick, Bhowmick2}.  Death scenario i.e HSS or IHSS state is only possible when $\epsilon_{1}\neq 0, \epsilon_{2}\neq 0$. In our study, for simplicity, we consider $\epsilon_1= \epsilon_2= \epsilon $. One of the solution of system (3) is (0, 0, 0, 0). The characteristic equation of coupled system (3) at origin is given by equation (2),
where $a_1=4\epsilon -2b,\;\; a_2=4\epsilon ^2+b ^2-6\epsilon b+2, \;\; a_3=4\epsilon-2b-4\epsilon ^2b+2b^ 2\epsilon$ and $a_4=1-2b\epsilon.$ Using Routh-Hurwitz criteria the origin i.e. HSS state is stable for $0.15\leq \epsilon \leq  1.66.$

\begin{figure} 
\centerline{\includegraphics[scale=0.6]{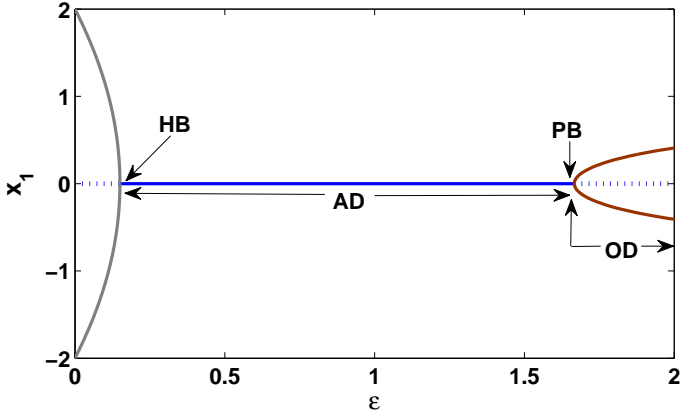}} 
\caption{Bidirectionally coupled Van der Pol oscillators. MATCONT plot shows a transition from HSS to IHSS.  HSS
appears via HB from oscillatory state at $\epsilon=0.15$ whereas IHSS occur via PB at $\epsilon=1.66$.}
\label{bidirectional_vdp}
\end{figure}

Numerical study of HSS-IHSS transition of  coupled Van der Pol oscillators is shown in Fig. \ref{bidirectional_vdp} where the bifurcation diagram
shows a HSS window (solid blue line) in a range of coupling strength and appears via HB. At the left side of the HSS window, a stable LC (solid gray line) exists where the trivial fixed point origin is unstable (dashed blue line). A transition from HSS to IHSS occurs at the right side of the HSS window via PB at $\epsilon=1.66$. Beyond the PB point, two distinctly new stable equilibrium points (solid brown line) are created as IHSS lines and coexist with the unstable fixed point origin (dashed blue line).\\
 \\
Thus, in another limit cycle system such as the VDP oscillator, the HSS-IHSS transition occurs via PB as usual irrespective of the type of coupling, typical diffusive or cyclic and the nature of parameter mismatch.
Next we focus on chaotic oscillators using similar coupling strategies (cyclic and diffusive) and a negative parameter mismatch in search of other bifurcation route if at all exists  besides the TB and the SNB under repulsive coupling \cite{Hens}.

\begin{figure}
 \centerline{\includegraphics[height=6cm,width=6.5cm]{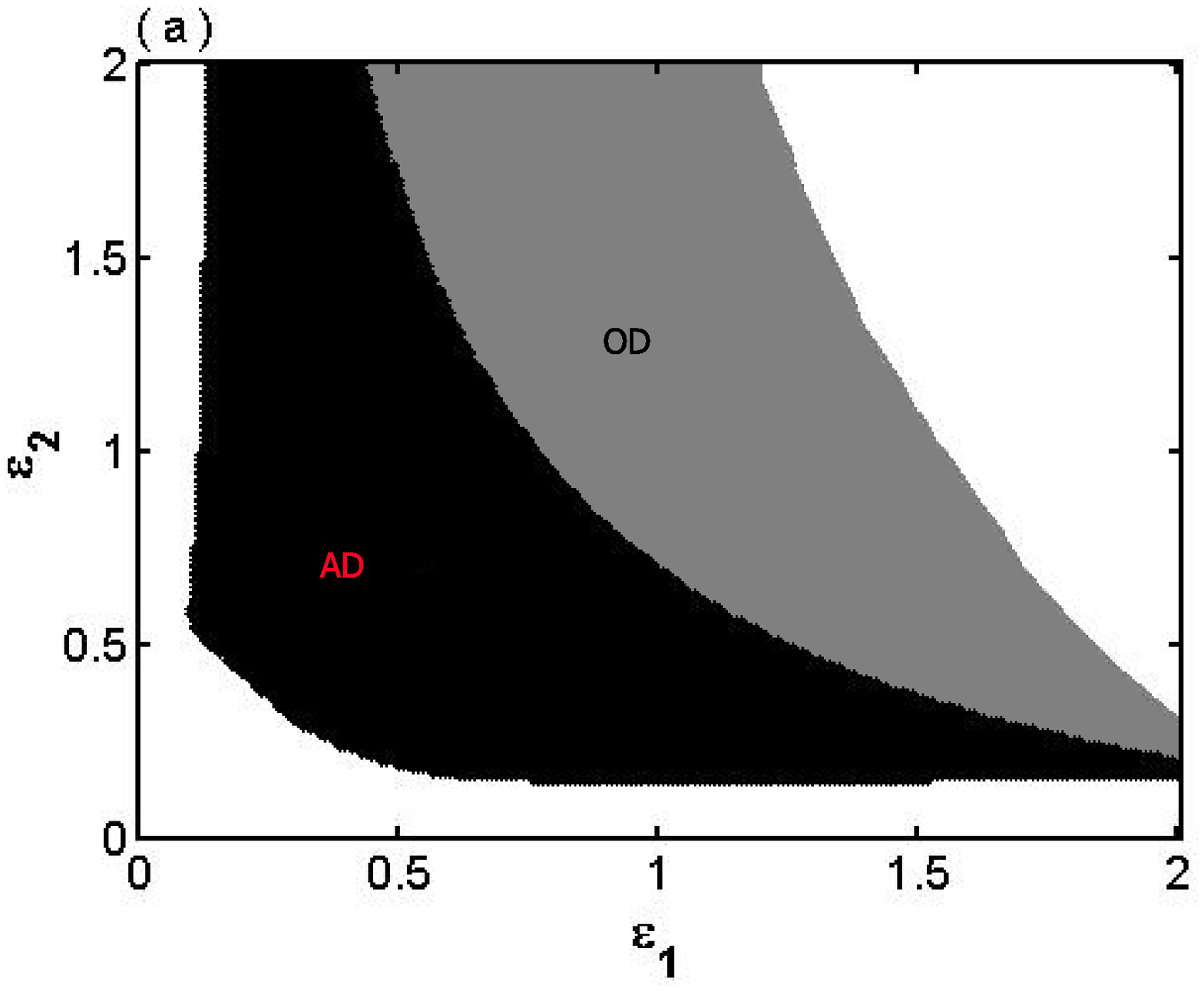}}
\centerline{\includegraphics[height=6cm,width=6.5cm]{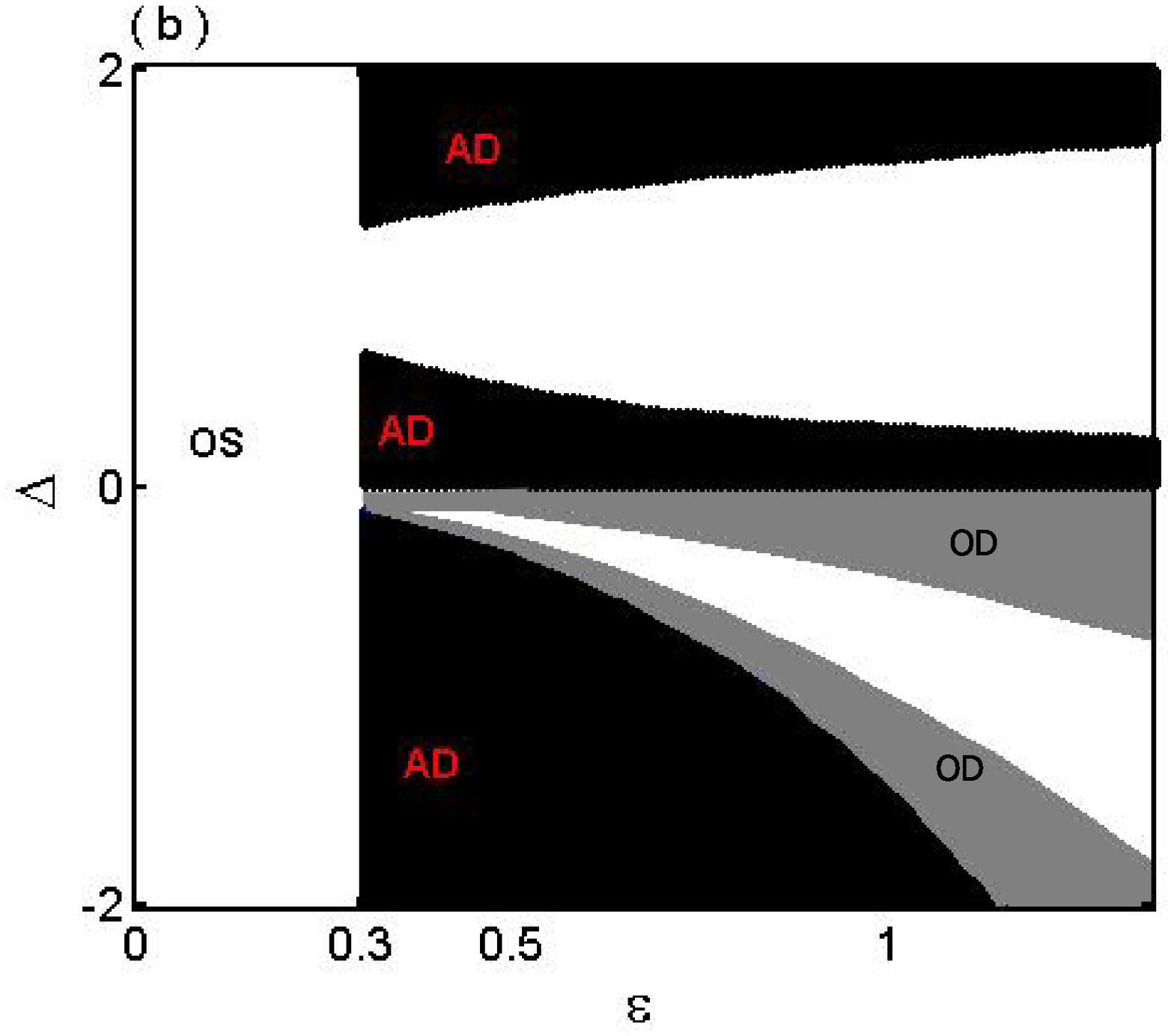}}
 \centerline{\includegraphics[height=6cm,width=6.5cm]{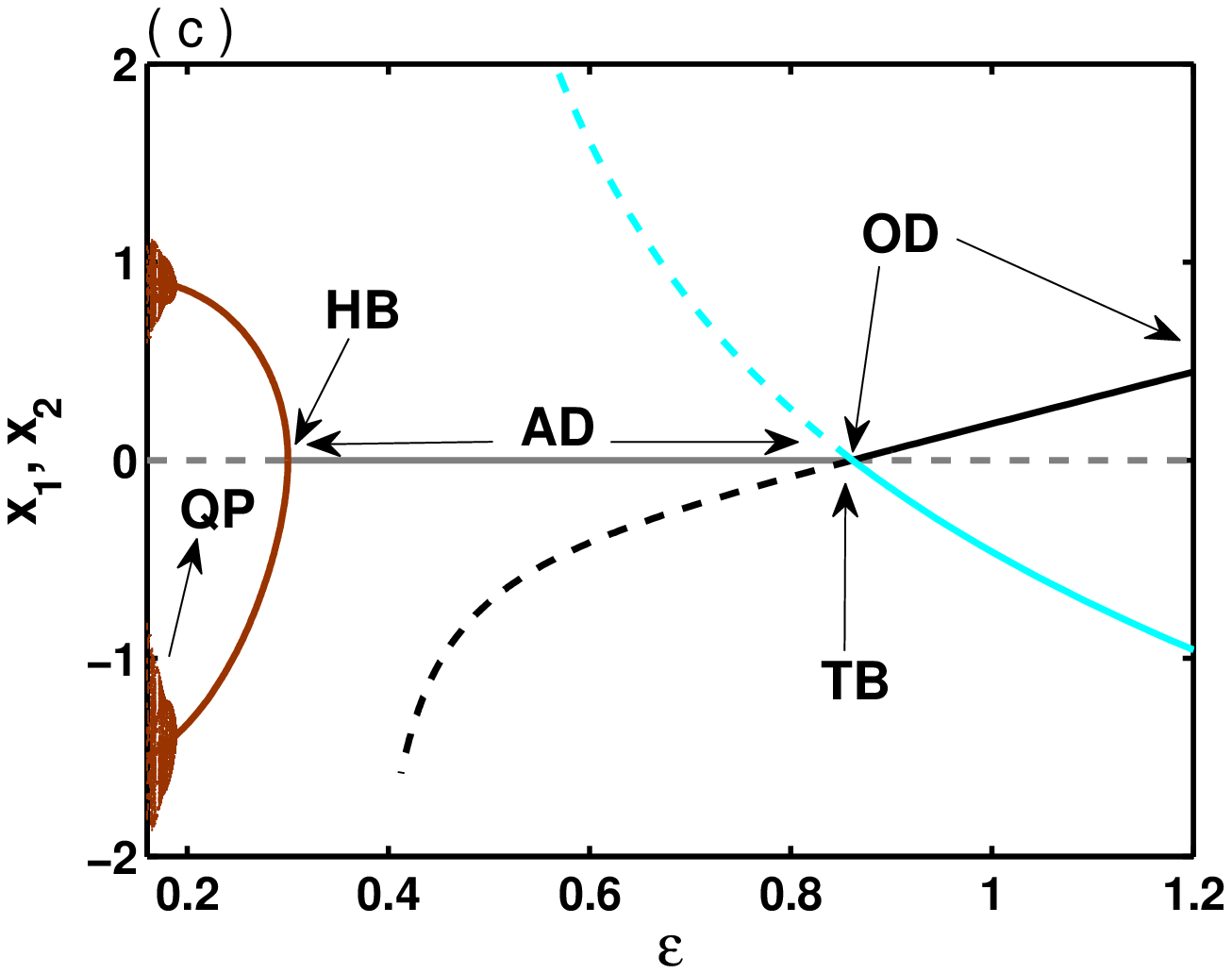}}
\caption{Coupled chaotic Sprott systems under cyclic coupling. 
The HSS and IHSS islands in (a) $\epsilon_1$-$\epsilon_2$ and (b) $\epsilon-\Delta$ plane. Black color represents the HSS (AD) region and grey lines are the boundaries of IHSS (OD) region. (c) Extremum of $x_1$ and $x_2$ is plotted with varying $\epsilon$ which corresponds to parameters along the   diagonal line $\epsilon_1= \epsilon_2$ in (a) and $\Delta =-1$ line in (b).  A transition from LC to AD is seen at $\epsilon=0.3$ followed by a transition to OD at $\epsilon=0.86$.}
\label{sprott_cyclic}
 \end{figure}

\par For numerical study,  we consider two chaotic Sprott oscillators under cyclic coupling appear as,
\begin{eqnarray}
&&\dot x_{1}=x_{1}y_{1}-\omega_1z_{1}+\epsilon_1(x_{2}-x_{1})\nonumber\\
&&\dot y_{1}=x_{1}-y_{1}\nonumber\\
&&\dot z_{1}=\omega_{1}x_{1}+az_{1}
\nonumber\\
&&\dot x_{2}=x_{2}y_{2}-\omega_{2}z_{2}
\nonumber\\
&&\dot y_{2}=x_{2}-y_{2}\nonumber\\
&&\dot z_{2}=\omega_{2}x_{2}+az_{2}+\epsilon_2(z_{1}-z_{2})
\end{eqnarray}
 The Sprott oscillator is chaotic in isolation for $a=0.3$; one oscillator sends a signal via $x$ variable and receives a feedback via $z$ variable which is expressed as cyclic coupling. 
The fixed points are given by the origin $(0, 0, 0, 0, 0, 0)$ and the $(x_1^*, y_1^*, z_1^*, x_2^*, y_2^*, z_2^*)$ where $y_1^*=x_1^*, z_1^*=-\frac{\omega_1x_1^*}{a}, x_2^*=\frac{1}{\epsilon_1}(\epsilon_1x_1^*-\frac{\omega_1^2x_1^*}{a}-x_1^{*2}),y_2^*=x_2^*, z_2^*=\frac{x_2^{*2}}{\omega_2}$ where $x_1^*$ is the real root of the cubic equation given by $\frac{\omega_2}{\epsilon_1}(\epsilon_1-\frac{\omega_1^2}{a}-x_1^*)+\frac{(a-\epsilon_2)x_1^*}{\omega_2\epsilon_1^2}(\epsilon_1-\frac{\omega_1^2}{a}-x_1^*)^2-\frac{\omega_1\epsilon_2}{a}=0$. 
 \par We induce a negative parameter mismatch $\Delta=\frac{\omega_2}{\omega_1}=-1$ as an asymmetry in the coupled system when the two oscillators are in counter-motion \cite{Bhowmick, Witkowski, Prasad,  Newby}. Assuming $\epsilon_1=\epsilon_2=\epsilon$, the cubic equation in $x_1^*$ is found to have a singularity at $\epsilon =a= 0.3$. For $\epsilon<0.3$ the equation has three non-zero real roots which are unstable and just after $\epsilon >0.3$ the equation has only one non-zero real root. This non-zero real root is unstable for $0.3<\epsilon <0.86$ whereas the HSS is stable. The eigenvalues of the Jacobian matrix of system (4) at origin are $\lambda_1=-1.0,  \lambda_2=-1.0$ and other eigenvalues are the roots of the equation given by (2)
where $a_1=2(\epsilon -a),\;\; a_2=\omega_1^2+ \omega_2^2-3a\epsilon +\epsilon ^2+a^2,\;\; a_3=(\epsilon-a)( \omega _1^2+\omega _2^2-a\epsilon),$ and $a_4=\omega_1^2 \omega_2^2+\epsilon ^2 \omega_1 \omega_2-a\epsilon  \omega_2^2. $ Using Routh-Hurwitz criteria, HSS is found to be stable in the range $0.3<\epsilon <0.86.$

  At $\epsilon \geq 0.86$, IHSS appears and HSS become unstable  via transcritical bifurcation (TB).  Further we expand the  regime of HSS and its transition to  IHSS by drawing   $\epsilon_1-\epsilon_2$ and  $\epsilon - \Delta$  phase space diagram in Fig. \ref{sprott_cyclic}(a) and \ref{sprott_cyclic}(b) respectively. For both the figures, HSS is shown in black color whereas IHSS is depicted in grey color.  The transition from HSS to IHSS is only possible when $\Delta <0$. Extreme values of $x_1$  and $x_2$ are plotted with $\epsilon$  as illustrated in Fig. \ref{sprott_cyclic}(c) using the software package MATCONT~\cite{Dhooge} for $\Delta =-1.0$. The chaotic dynamics becomes period-1 for weak coupling via quasiperiodic (QP) route and then the origin becomes stable as a HSS state via HB. Finally, the equilibrium origin loses its stability (from solid gray line to grey dotted line) and the unstable non-zero equilibrium points become stable as indicated by solid black lines and solid green lines. The HSS transits to IHSS via a TB at a larger coupling strength. A similar HSS-IHSS transition has also been observed in coupled chaotic oscillators under repulsive interaction~\cite{Hens, Hens2}. Note that, the white regimes in the right side of Fig. \ref{sprott_cyclic}(a) and between two OD islands in Fig. \ref{sprott_cyclic}(b) are unstable. We have checked that OD island ends via HB and goes to  the unstable regime. We focus here only at the emergence of  HSS and its transition to IHSS.
\begin{figure}[ht]
\centerline{
\includegraphics[scale=0.6]{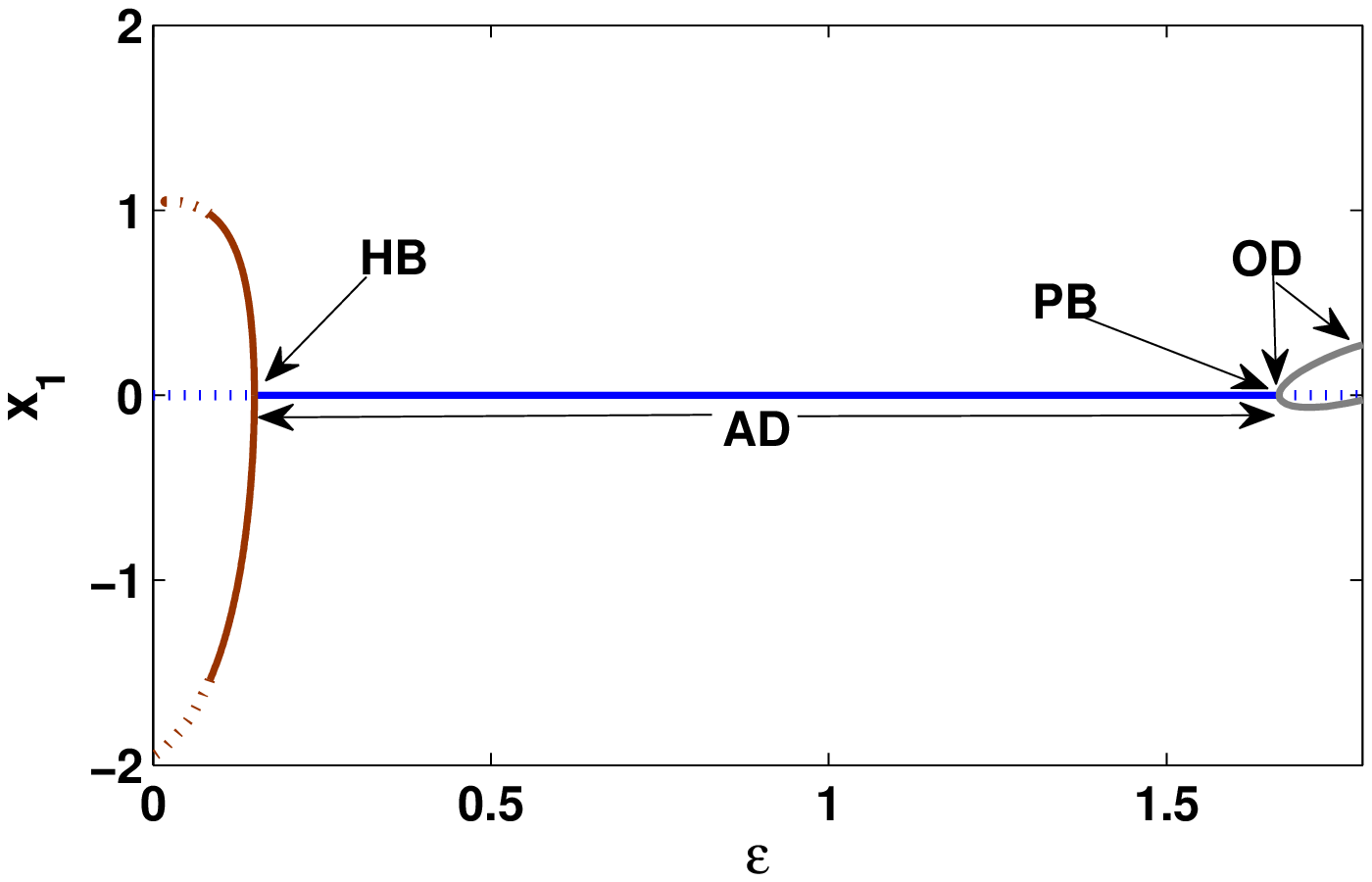}}
\caption{Coupled chaotic Sprott systems under diffusive coupling. Bifurcation diagram shows a transition of HSS (AD) to IHSS (OD) with varying coupling strength $\epsilon$. Extrema of $x_1$  has been plotted against $\epsilon$. Solid brown line represents Limit cycle oscillator where dotted blue line represents unstable region of LC. Solid blue line signifies HSS which occurs via HB and solid grey line in the right side represent two stable branch appears through PB which coexists with unstable origin.}
\label{Sprott_bifur}
\end{figure}

\par Finally, we search the transition scenario in coupled Sprott oscillators under the conventional diffusive bidirectional coupling. Two coupled Sprott systems under diffusive bidirectional coupling are given by
\begin{eqnarray}
&&\dot x_{1}=x_{1}y_{1}-\omega_1z_{1}+\epsilon_1(x_{2}-x_{1})\nonumber\\
&&\dot y_{1}=x_{1}-y_{1}\nonumber\\
&&\dot z_{1}=\omega_{1}x_{1}+az_{1}+\epsilon_2(z_{2}-z_{1})\nonumber\\
&&\dot x_{2}=x_{2}y_{2}-\omega_{2}z_{2}+\epsilon_1(x_{1}-x_{2})\nonumber\\
&&\dot y_{2}=x_{2}-y_{2}\nonumber\\
&&\dot z_{2}=\omega_{2}x_{2}+az_{2}+\epsilon_2(z_{1}-z_{2})
\end{eqnarray}
Clearly, the coupled system has a equilibrium point at origin. For simplicity, the coupling strengths  are  taken identical $\epsilon_1=\epsilon_2=\epsilon$. We draw a bifurcation diagram with varying the coupling strength $\epsilon$ in Fig. \ref{Sprott_bifur} using the MATCONT software.

 Figure \ref{Sprott_bifur}  shows a  transition from the HSS to the IHSS. For lower coupling, an unstable equilibrium origin (dashed line) coexists with stable LC
(solid brown line). Increasing the coupling strength HSS (solid blue line) originates via a HB.  A further increase of  coupling strength induces a transition to two separate branches of IHSS states via PB as usually found for all other forms of diffusive coupling \cite {Volkov, Kurths-pre, Banerjee}.

In summary, we numerically and analytically investigate a transition from HSS to IHSS for coupled  limit cycle and also for coupled  chaotic oscillators when their parameters are detuned. We use a VDP oscillator  as paradigmatic model for limit cycle and  sprott oscillator as chaotic model. To investigate the transition phenomena, we  introduce  cyclic coupling as well as  conventional diffusive coupling  when the oscillators have  negative parameter mismatch. We notice, conjugate effect of cyclic/diffusive coupling with detuned negative mismatch breaks the symmetry of the coupled system. In the limit cycle oscillators such as the VDP model under symmetric cyclic coupling or diffusive coupling, the HSS to IHSS transition appears via PB, as usually true for other limit cycle systems, so far investigated, irrespective of the coupling forms. On the other hand, for a chaotic Sprott system under cyclic coupling, the transition  occurred via TB as reported earlier for other chaotic systems under a combination of diffusive and repulsive mean field interaction. However, we find that the transition from HSS to IHSS in chaotic Sprott system follows a PB when two of them are coupled with the conventional diffusive bidirectional coupling. A transition of a homogeneous medium to stable patterns was first noticed by Turing \cite{Turing}. Very recently this becomes sudden issue of interest when a similar transition from a stable HSS to multiple IHSS state was reported \cite{Volkov} in Landau-Stuart (LS) limit cycle system. While most of the studies focused on the LS system only using different coupling forms \cite{Volkov, Kurths-pre,  Banerjee1}, none of them gives attention to chaotic system if such a transition could appear except by using a repulsive mean-field interaction \cite{Hens, Hens2}. Under cyclic or diffusive coupling in different negatively detuned mismatch oscillators  we examine the phenomenon of HSS to IHSS  with strong numerical and analytical evidence.\\
\\
\par {{\bf Acknowledgments:} {\it Authors would like to thank an anonymous referee for constructive criticisms and suggestions which has helped in improving the manuscript in its present form. They also thank S. K. Dana for useful discussions.}}

{}

\end{document}